\documentclass[a4paper,11pt]{article}
\usepackage{jheppub}
\usepackage{epsf}
\usepackage{undertilde}
\usepackage{amssymb}
\usepackage{amsmath}
\def\centerbox#1#2{\centerline{\epsfxsize=#1\textwidth \epsfbox{#2}}}
\def\p{\mathbf{p}}
\def\x{\mathbf{x}}
\def\Eq#1{Eq.~(\ref{#1})}
\def\OO{\mathcal{O}}
\def\Tr{\,\mathrm{Tr}\,}
\def\PhiL{\Phi_{_{\rm L}}}
\def\df{d_F}
\def\cf{C_F}

\def\g{g_{\rm 3d}}
\def\da{d_A}
\def\ca{C_A}
\def\crr{C_R}
\def\mD{m_{_{\!\mathrm{D}}}}
\def\be{\begin{equation}}
\def\ee{\end{equation}}
\def\bea{\begin{eqnarray}}
\def\eea{\end{eqnarray}}
\def\mybox{\begin{picture}(8,7)
  \thicklines\multiput(0,-0.3)(6.4,0){2}{\line(0,1){7}}
  \multiput(-0.3,0)(0,6.4){2}{\line(1,0){7}} \end{picture}}

\title{Renormalization of Null Wilson Lines in EQCD}

\author[a]{Michela D'Onofrio}
\author[b,c]{Aleksi Kurkela}
\author[b]{Guy D.\ Moore}

\affiliation[a]{Department of Physics and Helsinki Institute of
  Physics, University of Helsinki, P.O.\ Box 64, FI-00014 Helsinki, Finland}
\affiliation[b]{McGill University, 3600 rue University,
  Montr\'eal QC H3A 2T8, Canada}
\affiliation[c]{CERN, Physics Department, 1211 Gen\`eve 23, Switzerland}

\emailAdd{michela.donofrio@helsinki.fi}
\emailAdd{a.k@cern.ch}
\emailAdd{guymoore@physics.mcgill.ca}

\abstract{Radiation and energy loss of a light, high-energy parton in a
perturbative Quark-Gluon Plasma is controlled by transverse momentum
exchange. The troublesome infrared contributions to transverse momentum
exchange can be computed on the lattice using dimensional
reduction to EQCD.  However a novel extended operator, the Null Wilson
Line of EQCD, is involved.  We compute the renormalization properties of
this object's lattice implementation to next-to-leading order, which should
facilitate its efficient calculation on the lattice.}

\begin{document}
\maketitle
\flushbottom

\section{Introduction}
\label{sec:intro}

The quark gluon plasma created in the laboratory
\cite{EXPT} appears to be strongly coupled, in that its description
requires quick thermalization and a small viscosity
\cite{HYDRO}.  This is presumably because the temperature, and therefore
the energy scale for most of the physics, is not far above the QCD
transition temperature, where the coupling is large.  But some of the
most important probes of the medium involve the interaction of very high
energy particles with the medium -- hard probes, the most prominent of
which is jet modification \cite{WHO}.  Even if the medium is strongly
coupled, the high energy of the jet introduces a large energy scale at
which QCD is weakly coupled.  Therefore there is hope that jet
modification can be understood perturbatively.  More likely, it can be
understood by treating the jet constituents and their evolution
(particularly, particle splitting) perturbatively, but treating the
interaction of jet constituents with the medium nonperturbatively.

The propagation of a sufficiently high energy excitation through the
medium can be described in terms of a null Wilson line, and the
transverse momentum exchange with the medium is related to the falloff
with distance of a parallel pair of such lines \cite{Wilson1,Wilson2}.
Specifically, the probability per length to exchange transverse momentum
$\Delta p_\perp$ is given by
\be
\label{defC}
\frac{(2\pi)^2 d\Gamma}{d^2 \Delta p_\perp \; dt} \equiv C(p_\perp)
\,,
\qquad
C(p_\perp) = \int d^2 x_\perp e^{i\p_\perp \cdot \x_\perp}
C(x_\perp) \,,
\ee
where $C(x_\perp)$ is determined by a Wilson loop with two null segments
of length $l$ and two transverse spatial components of length
$x_\perp$:
\bea
\label{CandW}
C(x_\perp) & = & \lim_{l\rightarrow \infty} - \frac{1}{\ell} \ln \Tr
W_{l\times x_\perp} \,,
\nonumber \\
W_{l\times x_\perp} & = & \langle U_{(0,0,0);(l,0,l)} U_{(l,0,l);(l,\x_\perp,l)}
  U_{(l,\x_\perp,l);(0,\x_\perp,0)} U_{(0,\x_\perp,0);(0,0,0)} \rangle \,,
\eea
where $U_{x^\mu;y^\mu}$ are straight Wilson lines from $x^\mu$ to $y^\mu$, and
the three entries are the time, transverse coordinate, and longitudinal
coordinate.  The Wilson loop is to be evaluated in the density matrix
describing the collision, which is presumably a thermal density matrix.
Knowledge of $C(p_\perp)$, or equivalently $C(x_\perp)$, is a key input
into models of medium induced jet energy loss and jet modification
\cite{Wilson2,Arnold:2008iy}.

The leading order perturbative form of $C(p_\perp)$ is fully known \cite{Arnold:2008vd}, and
for momentum transfer of order of the temperature or higher $p_\perp \gtrsim T$, the corrections to the leading order result are suppressed by $\mathcal{O}(g^2)$. 
For a soft momentum transfer $p_\perp \sim g T$, however, the introduction
of a soft scale forces one to use resummed perturbation theory, 
and the Next-to-Leading order correction arises already at the $\mathcal{O}(g^3)$-order, making 
the physics of soft momentum transfers significantly more complicated.

However, in a remarkable paper \cite{Simon}, Caron-Huot has shown that for soft momentum transfers, to NLO, the Wilson loop
$W_{l\times x_\perp}$ above can be replaced by a Wilson loop in the much
simpler theory of EQCD, that is, Quantum Chromodynamics dimensionally
reduced to three Euclidean dimensions, with the $A^0$ field converted into
an adjoint scalar field which we will call $\Phi$ (roughly speaking
$\g\Phi = i A^0$ and $\g^2 \sim g^2 T$).  Specifically,
\be
\label{defW}
W_{\ell\times x_\perp} \rightarrow \tilde U_{(0,0);(0,l)}
  U_{(0,l);(\x_\perp,l)} \tilde U_{(\x_\perp,l);(\x_\perp,0)}
  U_{(\x_\perp,0);(0,0)} \,.
\ee
There is now no time coordinate, only the transverse and $z$
coordinates.  The complication is that the Wilson lines which replace
the null lines in the 4-D version of $W$ are modified, still containing
the descendant of the $A^0$ field, which enters in the definition of
$\tilde{U}$:
\be
\label{utilde}
\tilde U_{(0,0);(0,l)} = \mathrm{Pexp} \int_0^l dz \;
T_a ( i A^a_z + g\Phi^a )\,.
\ee
The representation matrices $T_a$ should be in the same representation
as the propagating particle, which we will label $R$ (typically the
fundamental or adjoint representation).
The relative factor of $\g$ is because we absorbed a $\g$ factor in
defining $\Phi$.  The relative phase -- $A_z$ enters with an $i$ and
$\Phi$ does not -- is because $\Phi$ is a Euclidean continuation of
$A^0$ and the $i$ factor is absorbed in the Wick rotation.
The overall sign is reversed in
$\tilde U_{(\x_\perp,l);(\x_\perp,0)}$.  We will call this modified
Wilson line the null Wilson line of EQCD.

Perturbation theory fails near the QCD crossover because the theory is
genuinely strongly coupled there.  But it
is possible that the failure of perturbation theory at a few times the
crossover temperature arises because the 3D theory is strongly coupled,
while the short-distance physics involved in dimensional reduction
is not \cite{Mikko}.  In this case, a nonperturbative treatment of the
3D theory may still give useful information about QCD at the highest
temperatures achieved in heavy ion collisions.  If true, then the
nonperturbative nature in the interaction of a jet parton with the
medium is captured by the EQCD value of $C(p_\perp)$, which can be
measured on the lattice.  With this motivation, there
has been an upswing in interest, recently, in studying the Wilson loop
and $C(x_\perp)$ in EQCD on the lattice \cite{PaneroRummukainen}.  The
relation between continuum thermal QCD and continuum EQCD is known to
high perturbative order \cite{Mikko,DimRed1,DimRed2,DimRed3}, and the matching
of the action, and some operators, between continuum and lattice EQCD is
known to order $\g^2 a$ \cite{Oa2}%
\footnote{\label{footdim}%
    In the 3-D theory the gauge coupling $g^2$ is dimensionful, carrying
    units of energy or inverse length; so $g^2 a$ is a dimensionless
    quantity.}%
\footnote{\label{footmass}%
    Actually the $\OO(g^2 a)$ matching between continuum and lattice
    EQCD is incomplete; the mass parameter of the $\Phi$ field is only
    known to two-loop order \cite{MikkoPhi}, which is $\OO(a^0)$.
    Improving this parameter to $\OO(g^2 a)$ will require a 3-loop
    calculation, though there is a way to determine the matching within
    numerical EQCD simulations, which we outline in Appendix
    \ref{App1}.}%
.  But the Wilson line in \Eq{utilde} is a new
operator and its lattice implementation has not been studied beyond the
tree level.  In practice it is challenging to make lattice studies
quantitatively reliable without a calculation of the $\OO(\g^2 a)$
renormalization of the null Wilson line operator. This is true even if
the lattice spacing is taken very small, if one is simultaneously
interested in $C(x_\perp)$ at short distances; as we will argue below,
the corrections arising from the Wilson operator scale as the
\textsl{larger} of $\g^2 a$ and $a/x_\perp$. Indeed, the first efforts to
numerically determine the $C(p_{\perp})$ by Panero, Rummukainen, and 
Sch\"afer \cite{PaneroRummukainen} show how it is challenging to make contact with perturbation theory at $p_\perp \gg g T$, corresponding to small spatial separations. Therefore we believe that a
study of $\OO(a)$ corrections to the null Wilson line operator are
essential to the success of this program.  We carry out this calculation
in the remainder of the paper.

In the next section we set up the problem, by writing the Lagrangian of
EQCD and an expression for the Wilson line in the continuum and the
lattice, highlighting what is needed in an NLO matching calculation.
The section also shows why the $\OO(a)$ correction becomes more
important at small $x_\perp$.  The body of the calculation appears
in Section \ref{Sec:details}, which explains how to handle lattice
diagrams with null Wilson lines, and tabulates the (Feynman gauge)
contribution of each relevant diagram.  We close with a brief discussion.

\section{Statement of the Problem}
\label{Sec:Lagrangian}

\subsection{Lattice and continuum action, Wilson line}
\label{subsec:lattcontin}

EQCD is the theory of a 3-dimensional SU($N$) gauge field $A^i$ with
field strength $F^{ij} \equiv F^{ij}_a T^a$, together with an adjoint
scalar $\Phi\equiv \Phi^a T_a$ [with $T_a$ the fundamental
representation group generators normalized such that
$\Tr T_a T_b = \delta_{ab}/2$].  Writing the path integral as
$\int \mathcal{D}[A,\Phi] \exp(-S_{\rm EQCD})$,
the most general super-renormalizable action%
\footnote{
The Lagrangian could in addition contain a $\Tr \Phi^3$ term, but
at zero baryon number chemical potential it is forbidden by the 
charge conjugation symmetry.
}
 in the continuum is
\be
\label{Scontin}
S_{\rm EQCD,c} = \int d^3 x \left( \frac{1}{2\g^2} \Tr F^{ij} F^{ij}
+ \Tr D^i \Phi D^i \Phi
+\mD^2 \Tr \Phi^2
+\lambda_1 (\Tr \Phi^2)^2 + \lambda_2 \Tr \Phi^4 \right) \,,
\ee
where we have not shown the counterterm which subtracts UV divergences
from the $\Tr \Phi^2$ term%
\footnote{For the exact form of the counterterm see, e.g., Eq.(2.8) of \cite{Kajantie:1997tt}.
}. 
The three dimensional theory corresponds to the dimensionally reduced 
four dimensional QCD along a \emph{matching curve}, specifying the values of 
the parameters of EQCD as a function of four-dimensional parameters: $g,T$, and $N$ and the number and masses of quark species  $N_f$ and $m_i$. For explicit expressions see for example Eq. (5.2)-(5.5) of \cite{Kajantie:2002wa}. For quark mass dependence, see \cite{Laine:2006cp}.

 It is customary to introduce dimensionless
versions of the mass and scalar coupling terms, by defining%
\footnote{\label{footx}%
Note that, for SU(2) or SU(3), the $\Tr \Phi^4$ and
$(\Tr \Phi^2)^2$ terms are not independent, as
$\Tr \Phi^4 = (\Tr \Phi^2)^2/2$ for these groups.  So in these cases one
of the scalar terms can be eliminated in favor of the other.}
\be
\label{xandy}
y \equiv \frac{\mD^2[\mu_{\overline{\rm MS}}=\g^2]}{\g^4} \,,
\qquad
x_1 = \frac{\lambda_1}{\g^2} \,, \qquad
x_2 = \frac{\lambda_2}{\g^2} \,.
\ee

The corresponding lattice theory, with lattice spacing $a$, is defined
in terms of the link matrices $U_i(x)=U_{x;x+a\hat{i}}$
and the lattice scalar field $\PhiL$.  The lattice action is
\bea
\label{Slatt}
S_{\rm EQCD,L} & = & \frac{2N}{Z_g \g^2 a}
              \sum_{x,i>j}\left(1 - \frac{1}{N} \Tr \mybox_{x,ij} \right)
\nonumber \\ && {}
  +2 Z_\Phi \sum_{x,i} \Tr  \Big( \PhiL^2(x)
  - \PhiL(x) U_i(x) \PhiL(x+a\hat i) U^\dagger_i(x) \Big)
\nonumber \\ && {}
+ \sum_x Z_{4} \left[ (x_1{+}\delta x_1) \Tr \PhiL^4
   + (x_2{+}\delta x_2) \left( \Tr \PhiL^2 \right)^2 \right]
   + Z_2 (y {+} \delta y) \Tr \PhiL^2 \,, \;\;
\\
\label{defbox}
\mybox_{x,ij} & \equiv & U_i(x) U_j(x+a\hat{i}) 
           U^\dagger_i(x+a\hat{j}) U^\dagger_j(x) \,,
\eea
and the lattice implementation of $\tilde U$ is%
\footnote{\label{footexp}%
    Actually the implementation shown here is not unique; for instance,
    one can also replace $\exp(Z \PhiL) \to Z_0 + Z_1 \PhiL$, avoiding
    the need to exponentiate.  But there are advantages to the
    exponential choice; for instance, $Z_0,Z_1$ are already nontrivial
    in an abelian theory, for which $Z$ takes its tree-level value.
    We will only consider the exponential choice here.}
\be
\tilde U_{(0,0);(0,na)} = \prod_{m=0}^{n-1} 
   \exp\Big( Z T_{R}^a\PhiL^a(ma\hat{z}) \Big) 
 U_{z,R}(ma\hat{z}) \,,
\label{lattW}
\ee
for a Wilson line in the $R$ representation.
Note that there is no factor of $i$ in $\exp(Z \PhiL)$, which is
not a unitary matrix.

The value of the scalar field wave function normalization $Z_\Phi$ is
actually a free choice in implementing the lattice theory, corresponding
to the normalization choice for the lattice scalar field.  For instance,
Panero, Rummukainen, and Sch\"afer \cite{PaneroRummukainen} choose
$Z_\Phi = \g^2 a Z_g$ (which they call $6/\beta$).  Another sensible
choice would be $Z_\Phi = 1/(\g^2 a Z_g)$, so the lattice spacing enters
the action as a common multiplicative factor.  We will not choose a
specific prescription in this paper.  Instead, we focus on the
combinations $Z_g$, $Z^2/Z_\Phi$, $Z_2/Z_\Phi$, and
$Z_{4}/Z_\Phi^2$, which are invariant under this
normalization freedom.  At tree level we would have $Z_g = 1$,
$Z^2/Z_\Phi = \g^2 a=Z_{4}/Z_\Phi^2$ and
$Z_2/Z_\Phi = \g^4 a^2$.  The coefficients
$Z_g$, $Z_2/Z_\Phi$, $Z_4/Z_\Phi^2$, $\delta x_{1,2}$ and $\delta y$
are already known.  For completeness we list their values in
Appendix \ref{App1}. Our goal is to determine the remaining unknown
parameter $Z^2/Z_\Phi$, which controls the renormalization of the null
Wilson line of EQCD.

\subsection{Sensitivity of Wilson loop to Renormalization}
\label{subsec:sensitive}

\begin{figure}[t]
\centerbox{0.5}{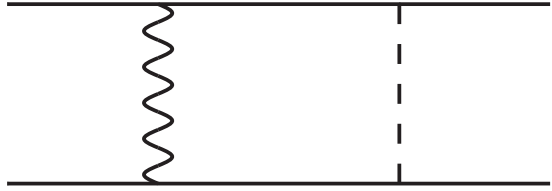}
\caption{Diagrams giving rise to the leading-order
  contribution to $C(x_\perp)$.  Wilson line (solid), showing the
  exchange between lines of an $A$ field (wiggly) or a $\Phi$ field
  (dashed)}
\label{fig:LO}
\end{figure}

Since we are interested in the $l$-dependence of $\Tr W$ when $l$ is
large, we can ignore contributions from the ends and corners of the
Wilson loop and focus on correlations between the long edges.
We are also only interested in the $x_\perp$ dependence of
$C(x_\perp)$, since any $x_\perp$-independent piece does not enter in
$C(p_\perp)$.  Therefore we need only consider diagrams with at least
one line connecting the null Wilson lines.  At lowest order there are
two, involving the exchange of an $A_z$ or a $\Phi$ line,
as illustrated in Figure \ref{fig:LO}.  Because the
$A_z$ fields attach with factors of $i,-i$ while the $\Phi$ fields
attach with factors of $1,-1$,
the contributions are of opposite sign.  In the continuum they are
\be
C_{\mathrm{LO}}(x_\perp) = \int_{-\infty}^\infty\!\!\! 
  dz \langle A_z(x_\perp,z) A_z(0)
 - \g^2 \Phi(x_\perp,z) \Phi(0) \rangle \quad \Rightarrow \quad
\frac{C(p_\perp)}{\crr}
= \frac{\g^2}{p_\perp^2} - \frac{\g^2}{p_\perp^2 + \mD^2}
\,,
\label{Cx_contin}
\ee
while on the lattice we find 
(defining, as usual $U_i(x) = \exp(1+ia A_i(x+a \hat{i} /2 ))$)
\be
C_{\mathrm{LO}}(x_\perp) = \frac{1}{a} \sum_n \langle
  a^2 A_z(x_\perp,na) A_z(0) - Z^2 \Phi(x_\perp,na) \Phi(0) \rangle
\;\; \Rightarrow \;\;
\frac{C(p_\perp)}{\crr}
 = \frac{Z_g g^2}{\tilde p_\perp^2}
   - \frac{Z^2/aZ_\Phi}{\tilde p_\perp^2+\mD^2} \,.
\label{Cx_latt}
\ee
Here $\tilde p_x^2 \equiv \sin^2(p_x a/2)/(a/2)^2$ is the lattice
propagator, and $\crr$ is the quadratic Casimir in the representation $R$
of the Wilson loop.

The important feature of \Eq{Cx_contin} is that the two terms
approximately cancel at large $p_\perp$, up to subleading
$\mD^2/p_\perp^4$ corrections.  The presence of the lattice propagator in
\Eq{Cx_latt} does not change this cancellation.  Of course this
cancellation does not persist at higher loop order; but because the
theory is super-renormalizable, each loop order gives weaker
large-$p_\perp$ behavior.  Indeed, at NLO the large $p_\perp$ behavior
is $\OO(\g^4/p_\perp^3)$ \cite{Simon}.

The problem is that the renormalization of $Z$ which is not taken into
account in a lattice calculation will spoil the cancellation in
\Eq{Cx_latt}, giving rise to uncanceled $1/p_\perp^2$ large-$p_\perp$
behavior -- specifically, a contribution of 
$(1 - Z^2_{\rm used}/Z^2_{\rm correct})\g^2/p_\perp^2$ -- in
\Eq{Cx_latt}.  Therefore
the short-distance or large-$p_\perp$ behavior is especially sensitive
to errors in the Wilson line renormalization constant $Z$.
Assuming $Z^2_{\rm used}/Z^2_{\rm correct} = 1 + \OO(\g^2 a)$, the
relative error is of order
$(\g^4 a/p_\perp^2)/(\g^4/p_\perp^3) \sim a p_\perp$,
corresponding to a $a/x_\perp$ relative error in $C(x_\perp)$.
%  ALEKSI:
%  [[This makes me wonder every time I look at the lattice data of PRS,
%  they do not see a 1/x behaviour in their data. Why is this?? Nor do
%  they even see a constant term. Is there a cancellation between the
%  $1/x_\perp$ and constant terms that make their data to look like its
%  approaching the origin from the positive side?]] 
%  GUY:
%  I think the key is that this is an estimate of the relative error.
%  We expect order-$x_\perp$ behavior, so this error looks like a
%  constant offset in $C(x_\perp)$.
Therefore the
need to renormalize the Wilson line operator increases at small
separation, scaling as the inverse separation of the Wilson lines in
lattice units.  For instance, if the Wilson lines are separated by
$N$ lattice spacings in the transverse direction, the $\OO(a)$
corrections are $\OO(1/N)$, no matter how small the lattice spacing may
be.  Finding the $\OO(a)$ correction to $Z$ will improve this behavior
to $1/N^2$, an important correction for realistic values $N\sim 5$.

\section{Calculation Details}
\label{Sec:details}

\subsection{Strategy}
\label{subsec:strategy}

The matching calculation consists of computing $C(p_\perp)/C_R$ at NLO
within continuum and lattice EQCD, and fixing the coefficients of the
lattice theory such that the calculations agree to all orders in $\g$ and $\lambda_i$ and up to the desired order in $a$, here $\mathcal{O}(a)$.  
As usual, once the coefficients are fixed at one order, the infrared behavior is
automatically the same at the next order, since the infrared behaviors
of the theories coincide by construction.  Then it is the difference in
the ultraviolet region of any loops which must be calculated.  As usual,
such behavior can be understood in terms of a renormalization of the
parameters of the theory appearing in diagrams of lower order.

\begin{figure}
\centerbox{0.95}{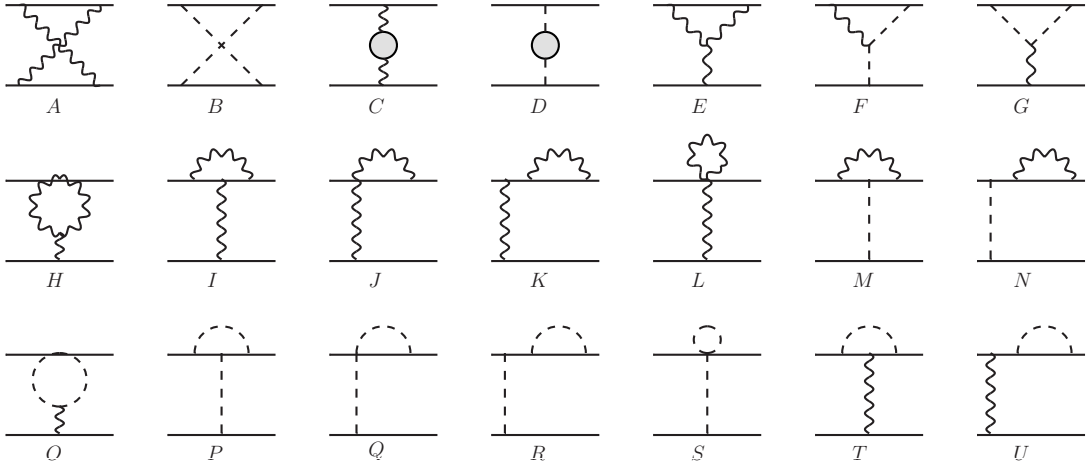}
\caption{Diagrams needed at next-to-leading order.  Solid lines are the
  Wilson lines, wiggly lines are $A$ fields, dashed lines are $\Phi$
  fields, blobs are self-energies.  Each diagram implicitly also
  represents the same diagram reflected right-left or top-bottom.}
\label{fig:NLO}
\end{figure}

Again we only need diagrams with at least one line running between the
null Wilson lines.
There are a number of NLO diagrams, see Figure \ref{fig:NLO}.
Fortunately, both propagators in diagrams $A$ and $B$ must be infrared
(since they connect spatially well-separated Wilson lines),
so they do not contribute to UV renormalization.  In Feynman
gauge (which we will use throughout), diagrams $E$, $H$, and $O$ are
zero.  Diagrams $J,L,Q,S$ have no continuum analog; they arise because
$U_z=\exp(iaA_z)$ and $\exp(Z\Phi)$ are nonlinear in $A_z$ and $\Phi$.
But the form of the lattice Wilson line, \Eq{lattW}, does not contain
anything which would introduce mixed $A_z,\Phi$ vertices on the Wilson
line, so there are no mixed-field analogs of diagrams $H,J,L,O,Q,S$.

Our strategy will be the following.  Since only the UV behavior of
diagrams is relevant, we can ignore $\mD$ and treat the propagators to
be
\be
\label{latt_props}
\langle A_z A_z(p) \rangle = \frac{Z_g \g^2}{\tilde p^2} \,,
\qquad
\langle \PhiL \PhiL(p) \rangle = \frac{a Z_\Phi^{-1}}{\tilde p^2}
\,.
\ee
In this case, for a soft momentum $p_\perp \ll 1/a$ running between the
Wilson lines, we extract all $1/p_\perp^2$ contributions, and choose the
value of $Z^2/Z_\Phi$ such that they cancel, as they do in the continuum
according to \Eq{Cx_contin}.

\subsection{Self-energies}
\label{subsec:self}

Diagrams $C$ and $D$ have been calculated \cite{Oa2}.  Indeed, diagram
$C$ makes up the principal contribution to $Z_g$ the gauge action
renormalization.  For this reason, we reproduce here the expression,
from \cite{Oa2}, for $Z_g$.  Ref.~\cite{Oa2} finds:
\begin{align}
\label{Zg}
Z_g^{-1} &= 1 + 2(Z_\Phi-1)+ 2(V_{A,L}-V_{A,c})+\frac{\left( \pi_{A,L}- \pi_{A,c} \right)}{p^2},
\end{align}
where $V_{A,L}$ and $V_{A,c}$ are the one loop contributions to the three-point gauge-scalar vertex on the lattice and in continuum, respectively, while $\pi_{A,L}$ and $\pi_{A,c}$ are the corresponding gauge self-energies. 
While $Z_g^{-1}$ itself is gauge invariant, the individual contribution of
each term to $Z_g^{-1}$ is gauge dependent. Nevertheless, we only
need the answer in Feynman gauge, in which they read:
\be
2(Z_\Phi - 1) + 2(V_{A,L}-V_{A,c}) = \frac{\g^2 a}{4} \left[ 8\ca \frac{\xi}{4\pi} \right]
\ee
\be
\frac{\left( \pi_{A,L}- \pi_{A,c} \right)}{p^2}  =   \frac{\g^2 a}{4}\left( \ca \left\{ \frac{1}{6} \frac{\Sigma}{4\pi}
               -\frac{2}{3} \frac{\xi}{4\pi} \right\} + \left[  \frac{4 \cf}{3} - \frac{\ca}{3} 
    + \frac{13\ca}{3}\frac{\xi}{4\pi} \right]\right). \label{eq:diffV}
\ee
We have written out color factors in terms
of the fundamental Casimir
$\cf$ and dimension $\df$, and the adjoint Casimir $\ca$ and dimension
$\da$:  in SU($N$) theory these are $\df = N$, $\cf = (N^2{-}1)/(2N)$,
$\ca = N$ and $\da = (N^2{-}1)$.
The constants appearing here are are
\be
\frac{\Sigma}{4\pi a} \equiv \int_{-\pi/a}^{\pi/a} \frac{d^3 p}{(2\pi)^3}
\frac{1}{\tilde p^2} \,, \qquad
\frac{\xi a}{4\pi} \equiv 
  \int_{-\pi/a}^{\pi/a} \frac{d^3 p}{(2\pi)^3} \frac{1}{(\tilde p^2)^2}
 - \int_{-\infty}^\infty \frac{d^3 p}{(2\pi)^3} \frac{1}{(p^2)^2}\label{eq:diffPi}
\ee
where in the latter expression we implicitly IR regulate both integrals
in the same way; numerically
$\xi = 0.152859324966101$ and $\Sigma = 3.17591153562522$.
These are the only constants we will need in the remainder of the
calculation.

The terms in curly brackets in Eq.~(\ref{eq:diffV}) arise from gauge self-energy diagrams with $\Phi$ running in the loops, and are, in fact, independent of the gauge parameter. The terms in square brackets arise from pure gauge self-energy 
diagrams%
\footnote{It may look strange that the pure-glue self-energy contains a
    rather large term proportional to $\cf$.  This is a tadpole-type
    contribution, which is dependent on the choice to implement the
    lattice link operators as fundamental-representation matrices.}
 and in a general gauge, would acquire a dependence on the gauge parameter. 

In Feynman gauge, the contribution from diagram C to $C(p_\perp)/C_R$ reads simply
\be
\mbox{Diagram $C$} = -\g^2 \frac{\left( \pi_{A,L}- \pi_{A,c} \right)}{(p_\perp^2)^2},
\ee
so that if we consider the sum of the self-energy diagram and the
leading-order $A_z$ exchange contribution, the self energy contributions
cancel and leave only the parts of $Z_g$ which arise from other sources:
\be
\label{diagramC}
\frac{Z_g \g^2}{p_\perp^2} + \mbox{Diagram $C$}
 = \frac{\g^2}{p_\perp^2} \left( 1 
   - \frac{\g^2 a \ca}{4}\: 8  \frac{\xi}{4\pi} \right) \,.
\ee

%Now we look at the distinct terms in \Eq{Zg} in more detail.
%The term in curly brackets arises from the $\Phi$ self-energy diagram.
%The division of the other terms between diagrams is gauge dependent; but
%we need the answer in Feynman gauge, in which the $8\ca \xi/4\pi$ arises
%from the $A\Phi\Phi$ vertex and $\Phi$ self-energy corrections, while the
%remaining terms come from the pure-glue self-energy.%
%\footnote{It may look strange that the pure-glue self-energy contains a
%    rather large term proportional to $\cf$.  This is a tadpole-type
%    contribution, which is dependent on the choice to implement the
%    lattice link operators as fundamental-representation matrices.}
%Therefore, when we consider the sum of the self-energy diagram and the
%leading-order $A_z$ exchange contribution, the self-energy contributions
%cancel and leave only the parts of $Z_g$ which arise from other sources:
%\be
%\label{diagramC}
%\frac{Z_g \g^2}{p_\perp^2} + \mbox{Diagram $C$}
% = \frac{g^2}{p_\perp^2} \left( 1 
%   - \frac{g^2 a \ca}{4}\: 8  \frac{\xi}{4\pi} \right) \,.
%\ee

The scalar self-energy is also computed in Ref.~\cite{Oa2}, where it is
responsible for the quantity called $Z_\Phi$ there.  Re-computing to
convert from Landau to Feynman gauge using the result for the
self-energy in Ref.~\cite{Oa1}, we find the sum of the tree level
and self-energy $\Phi$ exchange diagrams to give
\be
\label{diagramD}
-\frac{\g^2 Z^2/aZ_\Phi}{\tilde p_\perp^2} + \mbox{Diagram $D$}
 = - \frac{Z^2/a Z_\Phi}{\tilde p_\perp^2}
 \left( 1 + \frac{\g^2 a \ca}{4} \left[ 8 \frac{\xi}{4\pi}
    + \frac{2}{3} \frac{\Sigma}{4\pi} \right] \right) \,.
\ee

\subsection{Vertex corrections}

Next consider diagrams $M,N$.  Here it is relevant that \Eq{CandW}
involves $\ln\Tr W$, not just its trace.  In an abelian theory the
Wilson loop is the exponential of all 1-propagator corrections%
\footnote{%
    Here it is essential that both the $A$ and $\Phi$ field attachments
    are implemented via exponentials in \Eq{lattW}.  In the
    implementation suggested in Footnote \ref{footexp}, exponentiation
    would fail for the $\PhiL$ field.}%
, which
means that the abelian parts of diagrams $I-N$ and $P-U$ are absorbed
when we take the log.  Only the nonabelian parts of these diagrams
contribute.  The group theory factor in diagram $N$ is
$T^a T^a T^b T^b = \crr^2$, the product of the group factors for each
line.  Therefore $N$ is abelian.  Diagram $M$ involves
$T^a T^b T^a T^b = \crr(\crr-\ca/2)$; the $\crr^2$ piece is the abelian
part, the $\crr\ca$ piece is the nonabelian part we need.  Label the
momenta on the $\PhiL$ and $A_z$ lines $p$ and $q$ respectively.
The $\PhiL$ line can attach anywhere on each Wilson line.  The sum over
locations on the lower Wilson line gives a factor $\ell/a$, which cancels
the $a$ in the propagator and gives the $\ell$ which should be canceled
in \Eq{CandW}.  The sum over the location of the upper Wilson line gives
a factor $\delta(a p_z)$ which ensures that $p$ is purely transverse;
the $\Phi$ line then gives rise to the $Z^2/aZ_\Phi / \tilde{p}_\perp^2$
term found in \Eq{Cx_latt}.

Next we sum over the attachment positions of the $A_z$ propagator.  It
is most convenient to consider the link matrix $U_z(x)$ to ``live'' at
the center of the link $x+a\hat{z}/2$.  In this case, for a line
momentum $q$, the sum over attachment locations gives
\be
\left( a\sum_{n=0}^\infty e^{iq_z (n+1/2) a} \right)
\left( a\sum_{m=0}^\infty e^{-iq_z(-m-1/2) a} \right) \,,
\ee
where the first (second) term sums over the attachment of the front
(back) vertex, relative to where the $\PhiL$ attaches.  The sum is
easily performed by splitting off the first term and shifting the
remaining terms:
\bea
\label{sum1}
a\sum_{n=0}^\infty e^{iq_z (n+1/2) a}
 &=& ae^{iq_z a/2} + ae^{iq_z a} \sum_{n=0}^\infty e^{iq_z (n+1/2) a}
\quad \Rightarrow \nonumber \\
a\sum_{n=0}^\infty e^{iq_z (n+1/2) a} & = & 
\frac{ae^{iq_z a/2}}{1-e^{iq_z a}} = \frac{ia}{2\sin(q_z a/2)}
\equiv \frac{i}{\tilde q_z} \,.
\eea
This term will always arise when summing over the location of an $A_z$
attachment which must be to the right of a $\PhiL$ attachment on the
Wilson line.  Therefore it makes sense to define it as the Feynman rule
for the propagator of the Wilson line between a $\PhiL$ and an $A_z$
attachment.  The corresponding continuum expression is
$i/ q_z$.

The group theoretical issues in treating diagrams $I,J,K,L$ are
similar.  Each involves a factor $\crr^2$ and a factor
$-\crr \ca$, with coefficient $-1/2$, $-1/4$, $0$, and $-1/6$
for $I$, $J$, $K$, and $L$ repectively.  The sum over the attachment
points in diagram $I$ is similar to that in diagram $M$, except that the
attachments must be separated an integer distance.  They therefore
involve the sum
\be
\label{sum2}
a \sum_{n=1}^\infty e^{iq_z n a} 
= e^{iq_z a} \left(1+a \sum_{n=1}^\infty e^{iq_z n a} \right)
= \frac{i e^{iq_z a/2}}{\tilde q_z}
= \frac{i \cos \frac{q_z a}{2}}{\tilde q_z} - \frac{a}{2}
\equiv \frac{i\utilde{q}_z}{\tilde q_z} - \frac{a}{2}  \,.
\ee
The sum of the nonabelian contributions from diagrams $I,J,K,L$ is therefore
\bea
\label{IJL}
I+J+K+L &=& -\ca \g^2 \frac{\crr \g^2}{\tilde p_\perp^2}
\int_{-\pi/a}^{\pi/a} \frac{d^3 q_\perp}{(2\pi)^3} \frac{1}{\tilde q^2}
\left( 
\left[ \frac{\utilde{q}_z}{\tilde{q}_z} + \frac{ia}{2} \right]^2_{I}
-ia\left[ \frac{\utilde{q}_z}{\tilde{q}_z} + \frac{ia}{2} \right]_{J}
-\left[\frac{a^2}{3}\right]_{L} \right)
\nonumber \\
& = & \frac{\crr \g^2}{\tilde p_\perp^2} (-\ca \g^2)
\int_{-\pi/a}^{\pi/a} \frac{d^3 q_\perp}{(2\pi)^3} \frac{1}{\tilde q^2}
\left( \frac{\utilde{q}_z^2}{\tilde{q}_z^2} - \frac{a^2}{12} \right)\,.
\eea
We can rewrite
\be
\label{work_utilde}
\utilde{q_z}^2 \equiv \cos^2 \frac{q_z a}{2}
 = 1 - \frac{a^2 \tilde{q}_z^2}{4} 
\ee
and therefore
\be
\label{IJL_final}
I+J+L = -\ca \g^2 \frac{\crr \g^2}{\tilde p_\perp^2}
\int_{-\pi/a}^{\pi/a} \frac{d^3 q_\perp}{(2\pi)^3} \frac{1}{\tilde q^2}
\left( \frac{1}{\tilde{q}_z^2} - \frac{a^2}{3} \right) \,.
\ee

The calculation of $P,Q,R,S$ proceeds similarly and the result is in
fact identical except for a factor of $(-Z^2/aZ_\Phi)^2$, which is 1 at
the level of precision needed in the current calculation.  On the other
hand, diagrams $M$ and $T$ each give
\be
\label{MandT}
M = T = +\ca \g^2 \frac{\crr \g^2}{\tilde p_\perp^2}
\int_{-\pi/a}^{\pi/a}  \frac{d^3 q_\perp}{(2\pi)^3} \frac{1}{\tilde q^2}
 \frac{1}{\tilde{q}_z^2} \,.
\ee
These cancel the like factors from $I,J,L,P,Q,S$, so that all vertex
correction diagrams add to
\be
\label{all_vertex}
\mbox{$I$--$N$ plus $P$--$U$} = \frac{\crr \g^2}{\tilde p_\perp^2}
 \ca \g^2 \int_{-\pi/a}^{\pi/a} \frac{d^3 q_\perp}{(2\pi)^3} 
  \frac{1}{\tilde q^2} \frac{2a^2}{3}
=  \frac{\crr \g^2}{\tilde p_\perp^2} \frac{\ca \g^2 a}{4}
       \frac{8}{3} \frac{\Sigma}{4\pi} \,.
\ee

\subsection{Y-diagrams}
\label{subsec:y}

Finally we consider diagrams $E,F,G$.  In Coulomb gauge the vertex
appearing in $E$ connects three $A_z$ propagators.  Labeling the lower
momentum $p$ and the upper left and right momenta $q$ and $p+q$, we find
that $p_z=0$ automatically.  Applying the vertex Feynman rule
(see Ref.~\cite{Rothe} page 383),
\be
\mbox{Diagram $E$ Vertex} = \utilde{q}_z \tilde{q}_z - \widetilde{2q_z} +
\tilde{q}_z \utilde{q}_z = 0 \,.
\ee
This is not surprising; for instance, there is certainly no $A_z^3$
continuum vertex, since $F_{ij}^2$ always involves two distinct labels
each appearing twice.  Diagrams $F$ and $G$ can be computed in a
straightforward way using the Feynman rules we have already found for
the attachment of lines to the Wilson line, and we find
\bea
F &=&  \frac{\crr \g^2}{\tilde p_\perp^2} (-2\ca \g^2)
 \int_{-\pi/a}^{\pi/a} \frac{d^3 q_\perp}{(2\pi)^3}
 \frac{1}{(\tilde q^2)^2} \tilde q_z
 \frac{1}{\tilde q_z}\,,
\label{F_is}
\\
G &=& \frac{\crr \g^2}{\tilde p_\perp^2} 2\ca \g^2
 \int_{-\pi/a}^{\pi/a} \frac{d^3 q_\perp}{(2\pi)^3}
 \frac{1}{(\tilde q^2)^2} (\tilde q_z \utilde q_z)
 \frac{\utilde{q_z}}{\tilde q_z} \,,
\label{G_is}
\\
\label{F_and_G} \hspace{-2ex}
F+G & = &  \frac{\crr \g^2}{\tilde p_\perp^2} 2\ca \g^2
 \int_{-\pi/a}^{\pi/a} \frac{d^3 q_\perp}{(2\pi)^3}
 \frac{1}{(\tilde q^2)^2} (1 - \utilde q_z^2)
 =  \frac{\crr \g^2}{\tilde p_\perp^2} \frac{\ca \g^2 a}{4}
   \left( - \frac{2}{3} \frac{\Sigma}{4\pi} \right) \!.
\eea

\subsection{Summing it up}

Summing the leading-order and subleading-order contributions of
form $1/p_\perp^2$, that is, \Eq{diagramC}, \Eq{diagramD},
\Eq{all_vertex}, and \Eq{F_and_G}, and requiring that the cancellation
of $1/p_\perp^2$ terms should occur, we find
\bea
0 & = & \frac{\g^2 \crr}{p_\perp^2} \left(
1 - \frac{Z^2}{Z_\Phi a} + \frac{\g^2 a \ca}{4}
\left\{ -8 \frac{\xi}{4\pi} - 8 \frac{\xi}{4\pi}
 - \frac{2}{3} \frac{\Sigma}{4\pi}
 + \frac{8}{3} \frac{\Sigma}{4\pi}
 - \frac{2}{3} \frac{\Sigma}{4\pi} \right\} \right)
\nonumber \\
\frac{Z^2}{Z_\Phi a} & = & 1 + \frac{\g^2 a \ca}{4}
\left( \frac{4}{3} \frac{\Sigma}{4\pi} - 16 \frac{\xi}{4\pi} \right)\,.
\label{main_result}
\eea
This constitutes our main result.

\section{Discussion}
\label{Sec:discussion}

We have found the 1-loop renormalization factor which should be included
in the lattice implementation of the EQCD null Wilson line.
Specifically, given the definition of the lattice action found in
\Eq{Slatt} and of the Wilson line operator in \Eq{lattW}, the ratio of
the normalization of the lattice scalar field $\PhiL$ appearing in the
Wilson line to its normalization in the action is given in
\Eq{main_result}, which we repeat for convenience:
\be
\frac{Z^2}{Z_\Phi a} = 1 + \frac{\g^2 a \ca}{4}
\left( \frac{4}{3} \frac{\Sigma}{4\pi} - 16 \frac{\xi}{4\pi} \right)\,.
\label{main_again}
\ee
Using this renormalization in the Wilson line will
facilitate faster and more accurate lattice calculations of the infrared
contribution to $\hat{q}$ and $C(p_\perp)$.  In particular, it
eliminates the last source of error (except for $\delta y$, see
Appendix \ref{App1}) which obstructs a quick and accurate continuum
extrapolation in the lattice determination of $C(x_\perp)$.

Structurally the most interesting feature of the calculation is the
tendency for diagrams to nearly cancel, when one sums over lines being
$A_z$ and $\PhiL = iA_0$.  This cancellation is broken in the UV because
the Wilson line is built out of $\PhiL$ fields appearing at integer
sites and $A_z$ fields appearing at half-integer links.  Therefore the
Wilson line propagator between two like-type fields differs in the UV
from that between opposite-type fields.

There are a few other physically interesting quantities which can be
computed with the same methodology as the calculation performed here.
It is pointed out in Ref.~\cite{NLOphotons} that $\hat{q}$ and its
semi-collinear analogue $\hat{q}(\delta E)$ can both be computed as
correlation functions of operators separated by adjoint null Wilson
lines.  The renormalization of the Wilson line found here can be adopted
in that problem, though a rather high-loop calculation of UV
contributions to the correlator will also be necessary.  We leave this
to be considered in future work.

\section*{Acknowledgements}

We would like to thank Jacopo Ghiglieri, Kari Rummukainen, and Urs Wiedemann for useful
discussions. This work was supported in part by the Natural Sciences and Engineering
Research Council of Canada. M.D.~was supported by the Magnus Ehrnrooth Foundation of Finland.

\appendix

\section{Scalar Mass and Self-Coupling Renormalization}
\label{App1}

Here we write the known 1- and 2-loop renormalizations of the scalar
self-couplings and mass in the notation of this paper; and we discuss
what would be involved in a full $\OO(a)$ (3-loop) determination of
$\delta y$, or how it could be avoided by using the lattice to measure
the requisite corrections.

The 1-loop renormalization $Z_g$ appears already in \Eq{Zg}.  The
remaining one-loop renormalizations can be found in Ref.~\cite{Oa2}, and
are:%
\footnote{$Z_2/\g^4 a^2 Z_\Phi$ was called $Z_m$ there, and the notation
  for $x_1$, $x_2$, as well as the division between $x$ and $Z_4$, was
  slightly different.\label{foot:notation}}
\bea
\frac{Z_4}{Z_\Phi^2} & = & \g^2 a \left( 1 - \g^2 a \ca
 \left[ \frac{1}{3} \frac{\Sigma}{4\pi} + 6 \frac{\xi}{4\pi} \right]
 \right)\,,
\\
\frac{Z_2}{Z_\Phi} & = & \g^4 a^2 \left( 1 + \g^2 a
 \left[ -\frac{\ca}{6} \frac{\Sigma}{4\pi}
       +\left(-3\ca + (N^2+1) x_1 + \frac{2N^2-3}{N} x_2 \right) 
       \frac{\xi}{4\pi} \right] \right)\,, \qquad
\\
\delta x_1 & = & \g^2 a \left( 3 + (N^2{+}7) x_1^2 
   + 2\frac{2N^2-3}{N} x_1 x_2 
   + \left( 3 + \frac{9}{N^2} \right) x_2^2 \right) \frac{\xi}{4\pi}\,,
\\
\delta x_2 & = & \g^2 a \left( N + 2 \frac{N^2-9}{N} x_2^2
  + 12 x_1 x_2 \right) \frac{\xi}{4\pi} \,,
\\
\delta y_{\rm 1\:loop} & = & \frac{-1}{\g^2 a}
  \left( 2 N + (N^2+1) x_1 + \frac{2N^2-3}{N} x_2  \right)
   \frac{\Sigma}{4\pi} \,.
\eea
Here we have departed from our previous pattern of writing everything in
terms of Casimirs and dimensions and have specialized to SU($N$) gauge
theory, because it is not obvious to us that the form of the quartic
interaction we have used can be considered without modification in more
general groups.

The renormalization of $y$ is known to two loops.  Besides the factor
$Z_2$ included above, one needs
\bea
-16\pi^2 \delta y_{\rm 2\:loop} & = & 
\left( (N^2+1) x_1 + \frac{2N^2-3}{N} x_2 \right)
\left( \frac{N \Sigma^2}{2} + N \Sigma \xi - 2 N \delta \right)
\nonumber \\ && {}
+ N^2 \left( \frac{7\Sigma^2}{8} - \frac{\Sigma \pi}{6}
+ \frac{31 \Sigma \xi}{6}
+ 2\kappa_1 - \kappa_4 - 4 \rho - 4 \delta \right)
\nonumber \\ & & {} +
\left( (N^2+1) x_1 (2N-2x_1) + \frac{2N^2-3}{N}x_2 (2N-4x_1)
\right. \nonumber \\ && \left. \hspace{12ex} {}
- \frac{N^4-6N^2+18}{N^2} x_2^2
\right) \left(\ln\frac{6}{\g^2 a}+\zeta-3\Sigma \xi \right)\,.
\eea
% This "Aleksi" version is missing the corrections which arise from
% using improved rather than tree level expressions for $g^2$ ($Z_g$),
% $x_1$ and $x_2$ in the action, see hep-lat/9709053 Eq.(45)
%\bea
%\delta y_{\rm 2\:loop}  &= & -\frac{1}{(4\pi)^2} \Bigg\{ \left[ 2N (N^2+1)x_1 + 2 (2N^2-3)x_2  \right]\left( \frac{1}{4}\Sigma^2 -\delta \right)  \\
%&&+ N^2\left[ \frac{5}{8}\Sigma^2 + \left(\frac{1}{2}-\frac{4}{3N^2}\right)\pi \Sigma - 4 \delta - (4\rho - 2\kappa_1+\kappa_4)\right]\\
%&&+ \left( \ln \frac{6}{a \g^2}+\zeta\right)\Bigg[ 2 N(N^2+1)x_1 + (2N^3-3)x_2 \\
%&& - 2(N^2+1)x_1^2
%- 4\frac{2N^2-3}{N} x_1 x_2 - \frac{(N^4-6N^2+18)}{N^2}x_2^2 \Bigg]\Bigg\}
%\eea
Here $\zeta$, $\delta$, and $4\rho-2\kappa_1+\kappa_4$ are additional
constants which are defined in Ref.~\cite{MikkoPhi}; specifically
$\zeta=0.08849$, $\delta=1.942130$, and
$4\rho-2\kappa_1+\kappa_4 = -1.968325$.
Note that the 1-loop contribution is parametrically $1/a$ and the
two-loop contribution is of order $a^0, \ln(\g^2 a)$.  Therefore, in a
complete $\OO(a)$ corrected study, one should also establish the
three-loop $\OO(a)$ correction to $\delta y$, which will be
parametrically of form
\be
\delta y_{3\mathrm{loop}} \sim \g^2 a \left( C_3 x^3 + C_2 x^2 + C_1 x
+ C_0 \right) \,.
\label{y_3loop}
\ee
(Really $C_3 x^3 = C_{30} x_1^3 x_0^0 + C_{21} x_1^2 x_2 + \ldots$
so there are 10 coefficients in all.)
The diagrammatic computation of this correction, and particularly of
$C_0$, appears
rather difficult.  However, there is an alternative to a diagrammatic
computation which could be attempted.  The key is that the SU($N$)
theory with $N>2$ has a phase transition at some $y_{\rm crit}(x_1,x_2)$
for all values of $x_1,x_2$.  A lattice study, at fixed $x$, can
find the critical value of $y$ at a given lattice spacing,
$y_{\rm crit}(a,x)$.  One then repeats for several values of $a$,
and examines the extrapolation to small $a$.
Since all other parameters are known up to $\OO(a^2)$ corrections,
the only source for $\OO(a)$ dependence in $y_{\rm crit}(a,x)$ is the
unknown ($x$-dependent) $\OO(a)$ correction to $\delta y$.  If the
lattice determination of $y_{\rm crit}$ is accurate enough to determine
the linear in $a$ behavior with precision, this constitutes an
evaluation of the terms in \Eq{y_3loop}, at a given value of $x$.  By
repeating for several $x$ values, one can reconstruct all terms.
In particular, one can determine $C_3$ by studying the
theory with the gauge fields switched off, and only scalar fields with
quartic interactions.  Since much more powerful algorithms exist to
study this theory (cluster, worm, multigrid), an accurate determination
of $C_3$ should be straightforward.

\end{document}